\documentclass[10pt,american]{IEEEtran}
\usepackage[T1]{fontenc}
\usepackage[latin9]{inputenc}
\usepackage{verbatim}
\usepackage{amsmath}
\usepackage{amssymb}
\usepackage{graphicx}
\usepackage{esint}

\makeatletter

\providecommand{\tabularnewline}{\\}

\usepackage{graphics}\usepackage{cite}
\usepackage{multirow}

\usepackage{babel}

\makeatother

\usepackage{babel}
\begin{document}

\title{On Frequency-Domain Implementation of Digital FIR Filters Using Overlap-Add
and Overlap-Save Techniques}
\author{Håkan Johansson, \textit{Senior Member}, \textit{IEEE}, and Oscar
Gustafsson, \textit{Senior Member}, \textit{IEEE}\\
 \thanks{The authors are with the Department of Electrical Engineering, Link\"oping University, Link\"oping, Sweden (e-mail:hakan.johansson@liu.se, oscar.gustafsson@liu.se).}}
\maketitle
\begin{abstract}
In this paper, new insights in frequency-domain implementations of
digital finite-length impulse response filtering (linear convolution)
using overlap-add and overlap-save techniques are provided. It is
shown that, in practical finite-wordlength implementations, the overall
system corresponds to a time-varying system that can be represented
in essentially two different ways. One way is to represent the system
with a distortion function and aliasing functions, which in this paper
is derived from multirate filter bank representations. The other way
is to use a periodically time-varying impulse-response representation
or, equivalently, a set of time-invariant impulse responses and the
corresponding frequency responses. The paper provides systematic derivations
and analyses of these representations along with filter impulse response
properties and design examples. The representations are particularly
useful when analyzing the effect of coefficient quantizations as well
as the use of shorter DFT lengths than theoretically required. A comprehensive
computational-complexity analysis is also provided, and accurate formulas
for estimating the optimal DFT lengths for given filter lengths are
derived. Using optimal DFT lengths, it is shown that the frequency-domain
implementations have lower computational complexities (multiplication
rates) than the corresponding time-domain implementations for filter
lengths that are shorter than those reported earlier in the literature.
In particular, for general (unsymmetric) filters, the frequency-domain
implementations are shown to be more efficient for all filter lengths.
This opens up for new considerations when comparing complexities of
different filter implementations.
\end{abstract}


\section{Introduction}

\label{section:introduction} \label{Introduction}

\IEEEPARstart{D}{igital} finite-length impulse response (FIR) filtering
(linear convolution) of an infinitely long (in practice very long)
input sequence can be efficiently implemented in the frequency domain
using overlap-add or overlap-save techniques \cite{Oppenheim_89,Mitra_06}.
These techniques make use of the discrete Fourier transform (DFT)
and its inverse (IDFT). In between the transforms, there is a diagonal
matrix whose diagonal elements are the filter's DFT coefficients,
hereafter also referred to as DFT filter coefficients. The DFT and
IDFT can be efficiently implemented using fast Fourier transform (FFT)
algorithms which is the reason for the overall efficiency. The basic
overlap-add and overlap-save principles are well known \cite{Oppenheim_89,Mitra_06},
but there are few publications that consider their fundamental implementation
properties. These techniques are however gaining an increasing interest,
in particular in applications requiring long and/or many filters.
Examples of such applications include equalization of chromatic dispersion
in optical communications \cite{Eghbali_14b,Martins2016,Kovalev_17,Bae2018,Wang_21,Bae_2021},
filter banks and channelizers with many channels \cite{Liu_2019,Nadal_2020,DeGaudenzi_2020,Kim_2020,Ruiz_2021},
filters with narrow transition bands (don't-care bands) \cite{Zheng_2021},
signal reconstruction/enhancement \cite{Pillai_2015,Wang_2022}, and
predistortion in multiple-input multiple-output (MIMO) systems \cite{Brihuega_2022}.

Most of the previous papers that utilize the overlap-save- or overlap-add-based
implementations focus on the applications and study the overall system
performance for different instances \cite{Kovalev_17,Liu_2019,Nadal_2020,Wang_21,Bae_2021,Ruiz_2021,Brihuega_2022}.
In this paper, the focus is instead on fundamental properties of the
overlap-add and overlap-save implementations. For these implementations,
as will be shown, the inevitable coefficient quantizations make the
overall system a time-varying system instead of the intended time-invariant
system. Hence, the analysis of coefficient quantization becomes more
complicated than for the time-domain implementation, where it suffices
to assess the frequency response of the filter with quantized coefficients
\cite{Jackson_96,Wanhammar_11}. A time-varying system, on the other
hand, cannot be characterized with a frequency response. Instead,
such a system can be characterized in two different ways. One way
is to represent it with a distortion function and aliasing functions
which can be derived from a multirate filter bank (MFB) representation
\cite{Vaidyanathan_93,Mehr_02}, or via block digital filter representation
which utilizes matrix-vector quantities \cite{Burel_04,Daher_2010}.
The other way is to use a periodically time-varying impulse-response
(PTVIR) representation which corresponds to a set of time-invariant
impulse responses and their respective frequency responses \cite{Vaidyanathan_93,Mehr_02}. 

The MFB and PTVIR representations make it possible to separate the
coefficient quantization analysis from the data quantization analysis,
which should be carried out separately \cite{Jackson_96,Wanhammar_11}.
The representations are also useful when analyzing the effect of using
shorter DFT lengths than theoretically required for a given impulse
response length and input signal block length, which also results
in a time-varying system. This occurs for example when designing the
filter using its DFT coefficients as design parameters, and when the
diagonal matrix between the DFT and IDFT is replaced with a more general
matrix, both options used as a means to reduce the overall approximation
error in the least-squares sense for given DFT and block lengths \cite{Burel_04,Daher_2010}.
In these generalized cases, the overall system is also referred to
as a block digital filter \cite{Burel_04}. Shorter DFT lengths also
occur when using zero padding in the frequency domain as a means to
carry out time-domain interpolation efficiently. An example of this
will be presented in Section \ref{sec:Design-Example} of this paper.

\subsection{Contributions \label{subsec:Contribution}}

The main contributions of this paper are as follows.
\begin{itemize}
\item Systematic derivations and analyses of the MFB and PTVIR representations
of the overlap-add and overlap-save frequency-domain implementations
are provided. Analysis of frequency-domain implementation of linear
convolution was also considered in \cite{Burel_04,Daher_2010}. However,
\cite{Burel_04,Daher_2010} expressed the overall system in terms
of the distortion and aliasing functions but did not explicitly express
the overall system in terms of the MFB and PTVIR representations considered
in this paper. Hence, this paper provides further insights for the
design, analyses, and understanding, as it derives representations
in terms of filters instead of matrix-vector quantities. It is noted
here that some parts of this contribution have been presented at a
conference \cite{Johansson_2015}, but only the basic principles of
the overlap-add technique. Here, it is extended to incorporate the
overlap-save technique and the additional contributions below.
\item Expressions for the impulse responses in the PTVIR representation
are derived and a detailed analysis of their lengths and relations
is provided. This has not been considered earlier in the literature.
The expressions hold for quantized coefficients as well, and are thus
useful when analyzing the effects of coefficient quantization which,
as mentioned before, should be carried out separately from the data
quantization analysis \cite{Jackson_96,Wanhammar_11}. As will be
shown, which is not obvious at first sight, the overlap-add and overlap-save
techniques have different impulse response properties when using quantized
coefficients (quantized DFT filter coefficients and complex exponentials
in the DFT/IDFT\footnote{In efficient FFT/IFFT implementations of the DFT/IDFT, the complex
exponentials in the DFT/IDFT are not explicitly quantized. Instead,
they are implicitly quantized throught the quantizations of the twiddle
factors in the FFT/IFFT architectures.}), as well as shorter DFT lengths than theoretically required. 
\item A comprehensive computational-complexity analysis is provided, and
the issue of selecting the optimal DFT length for a given filter length
is addressed. Based on those results, we derive accurate formulas
for estimating the optimal DFT lengths, which have not been reported
before and differ from other works where optimal design refers to
optimal overall filtering performance for fixed DFT and filter lengths
\cite{Burel_04,Daher_2010}. It will also be shown that, using optimal
DFT lengths, that minimize the computational complexities (multiplication
rates), the frequency-domain implementations become more efficient
than the corresponding time-domain implementations for filter lengths
that are shorter than those reported earlier in the literature \cite{Oppenheim_89,Lyons_96,Ishihara_2011}.
In particular, for general (unsymmetric) filters, the frequency-domain
implementations are shown to be more efficient for all filter lengths.
This result opens up for new considerations when comparing complexities
of different filter implementation options.
\end{itemize}

\subsection{Outline and Notations}

Following this introduction, Section \ref{sec:Overlap-Add-Overlap-Save-Techniques}
recapitulates the overlap-add and overlap-save techniques. Sections
\ref{sec:Filter-Bank-Representation} and \ref{sec:Time-Varying-System-Representati}
derive the MFB and PTVIR representations, respectively. Section \ref{sec:Impulse-Response-Properties}
analyzes the impulse response lengths and relations between the impulse
responses in the PTVIR representation whereas Sections \ref{sec:Design-Example}
and \ref{sec:Implementation-Complexity} provide design examples and
computational-complexity analysis, respectively. Finally, Section
\ref{sec:Conclusion} concludes the paper.

Throughout this paper, a sequence (discrete-time signal) is denoted
as $x(n)$. The Fourier transform of $x(n)$ is defined by
\begin{equation}
X(e^{j\omega})=\sum_{n=-\infty}^{\infty}x(n)e^{-j\omega n},\label{eq:Fourier transform}
\end{equation}
with $\omega$ {[}rad{]} being the frequency variable (angle), and
the inverse Fourier transform is given by
\begin{equation}
x(n)=\frac{1}{2\pi}\int_{-\pi}^{\pi}X(e^{j\omega})e^{j\omega n}d\omega.
\end{equation}
The $z$-transform of $x(n)$, $X(z)$, is obtained from \eqref{eq:Fourier transform}
by replacing $e^{j\omega}$ with the complex variable $z$. Further,
the $N$-point DFT of a length-$N$ sequence $x(n),\,n=0,1,\ldots,N-1$,
is defined by
\begin{equation}
X(k)=\sum_{n=0}^{N-1}x(n)e^{-j2\pi kn/N},\quad k=0,1,\ldots,N-1,
\end{equation}
whereas the IDFT is given by
\begin{equation}
x(n)=\frac{1}{N}\sum_{k=0}^{N-1}X(k)e^{j2\pi kn/N},\quad n=0,1,\ldots,N-1.
\end{equation}
We refer to $X(k)$ as the DFT coefficients of $x(n)$. For an impulse
response $h(n)$ of a filter, we refer to $H(k)$ as the DFT filter
coefficients. 

\section{Overlap-Add and Overlap-Save Techniques\label{sec:Overlap-Add-Overlap-Save-Techniques}}

The point of departure is that we are to implement a digital FIR filter
with the impulse response $h(n)$ of length $L$ (and thus having
a filter order of $L-1$), for an input sequence $x(n)$ generating
an output sequence $y(n)$. This corresponds to linear convolution
according to
\begin{equation}
y(n)=\sum_{p=0}^{L-1}h(p)x(n-p).\label{eq:convolution}
\end{equation}
For convenience in the equations that follow, we have here assumed
that the input $x(n)$ is zero for negative values of $n$. The linear
convolution can be implemented in the frequency domain using the overlap-add
and overlap-save methods as detailed below. Both methods utilize a
zero-padded impulse response sequence of length\footnote{The expressions and properties to be derived in Sections \ref{sec:Filter-Bank-Representation}--\ref{sec:Impulse-Response-Properties}
hold for all $N\geq L$. However, when $N<L+M-1$, i.e. the DFT length
is too short, the linear convolution is not properly implemented in
the frequency domain and the expressions and properties can then be
used to asses the errors that are introduced, as demonstrated in Example
2 in Section \ref{sec:Design-Example}.}
\begin{equation}
N=L+M-1,
\end{equation}
according to
\begin{equation}
h_{z}(n)=\begin{cases}
h(n), & n=0,1,\ldots,L-1,\\
0, & n=L,L+1,\ldots,N-1.
\end{cases}\label{eq:h(n)_zero-padded}
\end{equation}
In the implementation, the $N$ DFT filter coefficients of $h_{z}(n)$,
say $H(k),\,k=0,1,\ldots,N-1$, will be used. They are given by
\begin{equation}
H(k)=\sum_{n=0}^{N-1}h_{z}(n)e^{-j2\pi nk/N}=\sum_{n=0}^{L-1}h(n)e^{-j2\pi nk/N}.\label{eq:H(k)}
\end{equation}
Further, $M$ denotes the length of the input segments (output segments)
in the overlap-add (overlap-save) methods.

\subsection{Overlap-Add Method}
\begin{figure}[t!]
	\centering \scalebox{0.8}{\includegraphics{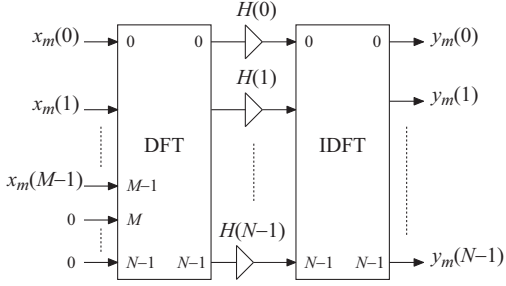}} \caption{Frequency-domain computation of the output segment $y_{m}(n)$ in
		the overlap-add technique.}
	\label{fig:OA-DFT-segment}
\end{figure}
In the overlap-add method \cite{Mitra_06}, the input sequence $x(n)$
is divided into adjacent input segments $x_{m}(n),\,m=0,1,2,\ldots,$
of length $M$. Then, each input segment is zero-padded to form a
sequence of length $N=L+M-1$ according to 
\begin{equation}
x_{m}(n)=\begin{cases}
x(n+mM), & n=0,1,\ldots,M-1,\\
0, & n=M,M+1,\ldots,N-1.
\end{cases}
\end{equation}
Also utilizing the zero-padded length-$N$ impulse response sequence
$h_{z}(n)$ in \eqref{eq:h(n)_zero-padded}, the output $y(n)$ can
then be computed as a sum of shifted and partially overlapping output
segments of length $N$ according to
\begin{equation}
y(n)=\sum_{m=0}^{\infty}y_{k}(n-mM),\label{eq:y(n)_sum_segments}
\end{equation}
where the output segments $y_{m}(n)$ are obtained from the convolution
\begin{equation}
y_{m}(n)=\sum_{p=0}^{N-1}h_{z}(p)x_{m}(n-p)=\sum_{p=0}^{L-1}h(p)x_{m}(n-p).
\end{equation}

\begin{figure}[t!]
\centering \scalebox{0.8}{\includegraphics{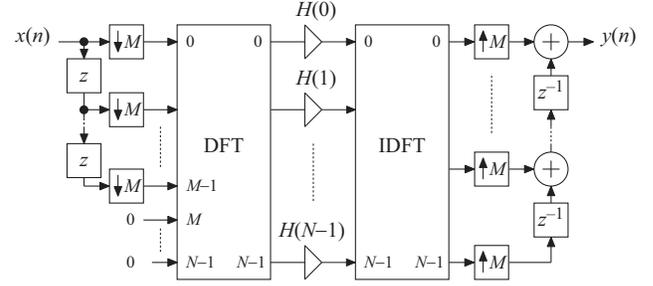}} \caption{Frequency-domain implementation using the overlap-add technique.}

\label{fig:OA-DFT} 
\end{figure}

Each output segment $y_{m}(n)$ can be computed by pointwise multiplying
$H(k)$ by $X_{m}(k)$, i.e., the length-$N$ DFT coefficients of
$x_{m}(n)$, and computing the length-$N$ IDFT of the so obtained
result. This is depicted in Fig. \ref{fig:OA-DFT-segment}. Since
the length of the output segments $y_{k}(n)$ is $N$, whereas each
of these segments is shifted $mM$ samples to form the output $y(n)$,
there is an overlap of $L-1$ samples between consecutive output segments.
For the overlapping time indices, the samples of the corresponding
output segments are consequently added to form the output samples.
For the remaining time indices, the output samples are taken directly
from the corresponding output segment. Utilizing upsamplers and downsamplers
\cite{Vaidyanathan_93}, the overlap-add method can be represented
by the structure\footnote{The noncausal (negative) delays, represented by $z$ in the structures
of Figs. \ref{fig:OA-DFT} and \ref{fig:OS-DFT}, are not explicitly
implemented. Together with the downsamplers, they are used to describe
how the input segments $x_{m}(n)$ can be generated from $x(n)$,
which is utilized in the multirate filter bank representation.} in Fig. \ref{fig:OA-DFT}.

\subsection{Overlap-Save Method}

In the overlap-save method \cite{Mitra_06}, the input sequence $x(n)$
is divided into overlapping input segments $x_{m}(n),\,m=0,1,2,\ldots,$
of length $N$ according to
\begin{equation}
x_{m}(n)=x(n+mM),\,n=0,1,\ldots,N-1.
\end{equation}
The output $y(n)$ can again be computed as a sum of output segments
$y_{m}(n)$ according to \eqref{eq:y(n)_sum_segments}. However, here,
$y_{m}(n)$ are length-$M$ segments and thus adjacent, not overlapping.
They are obtained as
\begin{equation}
y_{m}(n)=y_{mC}(n+L-1),\,n=0,1,\ldots,M-1,
\end{equation}
where each $y_{mC}(n)$ is the length-$N$ output of the circular
convolution between $h_{z}(p)$ and $x_{m}(n)$, as given by \cite{Mitra_06}
\begin{equation}
y_{mC}(n)=\sum_{p=0}^{N-1}h_{z}(p)x_{m}(n-p\textnormal{ mod }N).
\end{equation}

\begin{figure}[t!]
\centering \scalebox{0.8}{\includegraphics{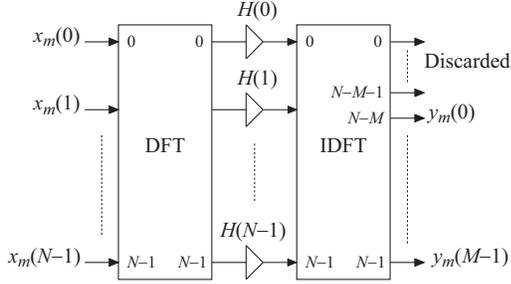}} \caption{Frequency-domain computation of the output segment $y_{m}(n)$ in
the overlap-save technique.}

\label{fig:OS-DFT-segment}
\end{figure}

Each output segment $y_{m}(n)$ can be computed by first pointwise
multiplying $H(k)$ by $X_{m}(k)$, then computing the length-$N$
IDFT of the so obtained result, and finally discarding the first $L-1$
values of the $N$ IDFT output values. This is illustrated in Fig.
\ref{fig:OS-DFT-segment}. An advantage of the overlap-save technique
is that the output segments do not overlap which means that the output
additions present in the overlap-add method are avoided. However,
there are also other implementation aspects to consider, which means
that the overlap-add technique may still be competitive as to the
overall implementation complexity. In particular, one needs to consider
the fact that the overlap-add technique uses a DFT with $L-1$ inputs
being zero, whereas the overlap-save technique uses an IDFT with $L-1$
outputs being unused. Hence, in both cases, some operations in the
DFT and IDFT may be removed. The exact amount of savings depend on
the architecture as well as the values of $L$, $M$, and $N$.

Finally, again utilizing upsamplers and downsamplers \cite{Vaidyanathan_93},
the overlap-save method can be represented by the structure in Fig.
\ref{fig:OS-DFT} where $H(k)$, $k=0,1,\ldots,N-1$, are again the
$N$ DFT coefficients of $h_{z}(n)$ given by \eqref{eq:H(k)}.

\section{Multirate Filter Bank Representation\label{sec:Filter-Bank-Representation}}

Using properties of DFT FBs \cite{Vaidyanathan_93}, the scheme in
Fig. \ref{fig:OA-DFT} can be equivalently represented by an $N$-channel
MFB, as depicted in Fig. \ref{fig:FB_representation}, with analysis
filters $G_{k}(z)$, $k=0,1,\ldots,N-1$, and synthesis filters $F_{k}(z)$,
$k=0,1,\ldots,N-1$, as described below for the two cases. It is stressed
that the MFB representation in Fig. \ref{fig:FB_representation} is
used for analysis purposes only. It should not be used for the implementation
of the overlap-add and overlap-save techniques, as its complexity
is higher than the complexities of the schemes in Figs. \ref{fig:OA-DFT}
and \ref{fig:OS-DFT}.

\subsection{Overlap-Add}

\begin{figure}[t!]
\centering \scalebox{0.8}{\includegraphics{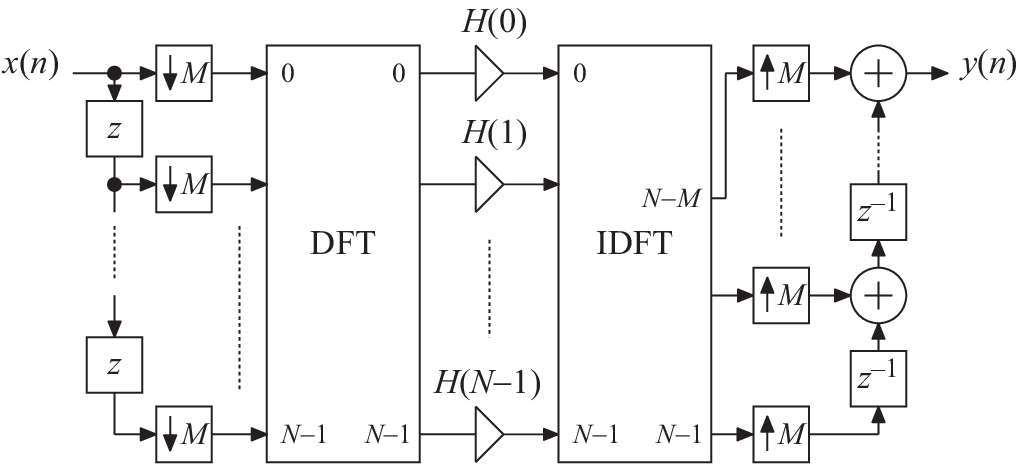}} \caption{Frequency-domain implementation using the overlap-save technique.}

\label{fig:OS-DFT}
\end{figure}

\begin{figure}[t!]
\centering \scalebox{0.8}{\includegraphics{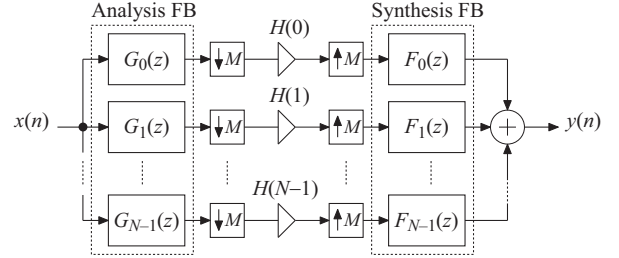}} \caption{MFB representation of the schemes in Figs. \ref{fig:OA-DFT} and \ref{fig:OS-DFT}.
It is used for analysis purposes only.}

\label{fig:FB_representation}
\end{figure}

In this case, the analysis and synthesis filters have length-$M$
and length-$N$ impulse responses, respectively, and are given by\footnote{Deriving $g_{k}(n)$ from the realizations in Figs. \ref{fig:OA-DFT}
and \ref{fig:OS-DFT}, one obtains noncausal analysis filters (due
to the use of $z$, see Footnote 1). To obtain the corresponding causal
filter impulse responses in \eqref{eq:gk_oa} and \eqref{eq:gk_os},
the noncausal filter impulse responses have been right-shifted $M-1$
and $N-1$ steps, respectively. This corresponds to replacing $n$
with $n-M+1$ and $n-N+1$, respectively. Further, since $e^{-j2\pi Nk/N}=1$
for all integers $k$, $N$ can be eliminated, which leaves only $n+1$
seen in \eqref{eq:gk_os}. Similarly, $n-M$ seen in \eqref{eq:fk_os}
for the overlap-save impulse responses $f_{k}(n)$, emanates from
a left-shift by $L-1=N-M$ samples due to the discard of $L-1$ IDFT
output samples.}
\begin{equation}
g_{k}(n)=e^{j2\pi(n-M+1)k/N},\quad n=0,1,\ldots,M-1,\label{eq:gk_oa}
\end{equation}
and
\begin{equation}
f_{k}(n)=\frac{1}{N}e^{j2\pi nk/N},\quad n=0,1,\ldots N-1.\label{eq:fk_oa}
\end{equation}
The corresponding frequency responses are
\begin{equation}
G_{k}(e^{j\omega})=\sum_{n=0}^{M-1}e^{j2\pi(n-M+1)k/N}e^{-j\omega n}\label{eq:FreqResp_OA_G}
\end{equation}
and
\begin{equation}
F_{k}(e^{j\omega})=\frac{1}{N}\sum_{n=0}^{N-1}e^{j2\pi nk/N}e^{-j\omega n}.\label{eq:FreqResp_OA_F}
\end{equation}

\subsection{Overlap-Save}

Here, the analysis and synthesis filters have length-$N$ and length-$M$
impulse responses, respectively, and are given by
\begin{equation}
g_{k}(n)=e^{j2\pi(n+1)k/N},\quad n=0,1,\ldots,N-1,\label{eq:gk_os}
\end{equation}
and
\begin{equation}
f_{k}(n)=\frac{1}{N}e^{j2\pi(n-M)k/N},\quad n=0,1,\ldots,M-1.\label{eq:fk_os}
\end{equation}
The corresponding frequency responses are
\begin{equation}
G_{k}(e^{j\omega})=\sum_{n=0}^{N-1}e^{j2\pi(n+1)k/N}e^{-j\omega n}\label{eq:FreqResp_OS_G}
\end{equation}
and
\begin{equation}
F_{k}(e^{j\omega})=\frac{1}{N}\sum_{n=0}^{M-1}e^{j2\pi(n-M)k/N}e^{-j\omega n}.\label{eq:FreqResp_OS_F}
\end{equation}

\subsection{Distortion and Aliasing Functions}

Based on the MFB representation in Fig. \ref{fig:FB_representation},
one can express the output Fourier transform $Y(e^{j\omega})$ as
\begin{equation}
Y(e^{j\omega})=\sum_{p=0}^{M-1}V_{p}(e^{j\omega})X\big(e^{j(\omega-2\pi p/M)}\big),
\end{equation}
where $V_{0}(e^{j\omega})$ is the distortion frequency response whereas
the remaining $V_{p}(e^{j\omega})$, $p=1,2,\ldots,M-1$, are aliasing
frequency responses. Using well-known input-output relations of MFBs
\cite{Vaidyanathan_93}, it follows that $V_{p}(e^{j\omega})$ are
given by
\begin{equation}
V_{p}(e^{j\omega})=\frac{1}{M}\sum_{k=0}^{N-1}H(k)G_{k}\big(e^{j(\omega-2\pi p/M)}\big)F_{k}(e^{j\omega})
\end{equation}
where $H(k)$ is given by \eqref{eq:H(k)} whereas $G_{k}(e^{j\omega})$
and $F_{k}(e^{j\omega})$ are the frequency responses of $g_{k}(n)$
and $f_{k}(n)$, as given by \eqref{eq:FreqResp_OA_G} and \eqref{eq:FreqResp_OA_F}
for overlap-add and by \eqref{eq:FreqResp_OS_G} and \eqref{eq:FreqResp_OS_F}
for overlap-save.

Using infinite-precision DFT and IDFT coefficients, we have\footnote{The frequency-domain implementations have an additional delay of $M-1$
samples due to the blockwise processing. For simplicity, this delay
is left out in the discussions in Sections \ref{sec:Filter-Bank-Representation}
and \ref{sec:Time-Varying-System-Representati}.} $V_{0}(e^{j\omega})=H(e^{j\omega})$, where $H(e^{j\omega})$ is
the frequency response of $h(n)$, i.e.,
\begin{equation}
H(e^{j\omega})=\sum_{n=0}^{L-1}h(n)e^{-j\omega n},
\end{equation}
whereas all aliasing terms are zero, i.e., $V_{p}(e^{j\omega})=0$
for $p=1,2,\ldots,M-1$. However, when the DFT coefficients and complex
exponentials in the DFT/IDFT are quantized (see Footnote 1), aliasing
will be introduced. This means that the frequency-domain implementation
of linear convolution corresponds to a weakly time-varying system
instead of the desired time-invariant system. The above representation
is a useful tool for analyzing the overall system performance when
quantizing the coefficients. One can thereby set requirements on $V_{0}(e^{j\omega})$
to approximate $H(e^{j\omega})$ and on $V_{p}(e^{j\omega})$, $p=1,2,\ldots,M-1$,
to approximate zero. Alternatively, depending on the application,
it may be better to use a PTVIR representation to assess the overall
performance.

\section{Periodically Time-Varying Impulse-Response Representation\label{sec:Time-Varying-System-Representati}}

An MFB with $M$-fold downsampling and upsampling, as in Fig. \ref{fig:FB_representation},
corresponds to an $M$-periodic linear system \cite{Mehr_02,Johansson_06}.
The output $y(n)$ of such a system, assuming an FIR system with impulse
response lengths $L_{n}$, is given by
\begin{equation}
y(n)=\sum_{q=0}^{L_{n}-1}h_{n}(q)x(n-q),
\end{equation}
where $h_{n}(q)=h_{n+M}(q)$ denotes the $M$-periodic impulse response
of the system. Due to the periodicity, such a system is completely
characterized by a set of $M$ impulse responses, $h_{n}(q)$, $n=0,1,\ldots,M-1$,
and thus by the $M$ corresponding frequency responses
\begin{equation}
H_{n}(e^{j\omega})=\sum_{q=0}^{L_{n}-1}h_{n}(q)e^{-j\omega q}.
\end{equation}
Using the inverse Fourier transform, the output can then alternatively
be written as
\begin{equation}
y(n)=\frac{1}{2\pi}\int_{-\pi}^{\pi}H_{n}(e^{j\omega})X(e^{j\omega})e^{j\omega n}d\omega.
\end{equation}
From the filtering point of view, this means that different output
samples are affected by different frequency responses, $H_{n}(e^{j\omega})$.
When $H_{n}(e^{j\omega})=H(e^{j\omega})$ for all $n$, the system
reduces to a regular linear and time-invariant filter with the frequency
response $H(e^{j\omega})$. In the frequency-domain implementation
of linear convolution, this is the desired result and corresponds
to $V_{0}(e^{j\omega})=H(e^{j\omega})$ and $V_{p}(e^{j\omega})=0$,
$p=1,2,\ldots,M-1$, in the MFB representation. Using the PTVIR representation,
the overall system performance is thus evaluated by studying $H_{n}(e^{j\omega})$,
all of which should approximate $H(e^{j\omega})$.

\begin{figure}[t!]
\centering \scalebox{0.8}{\includegraphics{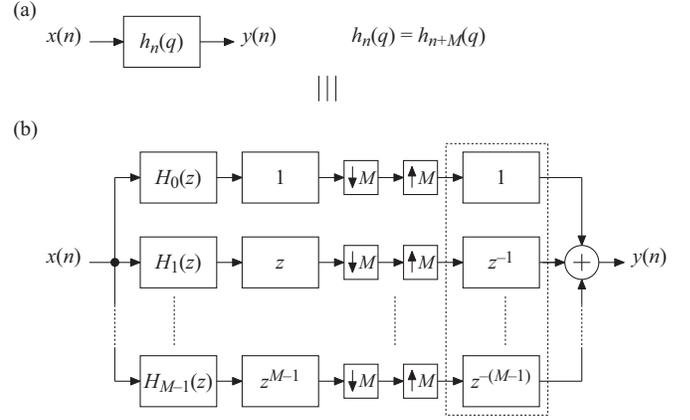}} \caption{PTVIR representations of the schemes in Figs. \ref{fig:OA-DFT} and
\ref{fig:OS-DFT}.}

\label{fig:TV_representation}
\end{figure}

The $M$ frequency responses $H_{n}(e^{j\omega})$ can be derived
from the analysis and synthesis filters in Fig. \ref{fig:FB_representation}.
To this end, it is first observed that the $M$-periodic impulse response
representation, depicted in Fig. \ref{fig:TV_representation}(a),
can be equivalently represented by the structure in Fig. \ref{fig:TV_representation}(b)
\cite{Johansson_06}. This structure can also be viewed as an MFB
representation, with trivial synthesis filters, but it should not
be confused with the general MFB representation in Fig. \ref{fig:FB_representation}
of Section \ref{sec:Filter-Bank-Representation}. Hence, regardless
of whether the representation in Fig. \ref{fig:TV_representation}(a)
or (b) is used, we still refer to it as the PTVIR representation.
Using polyphase decomposition, and properties of downsamplers and
upsamplers \cite{Vaidyanathan_93}, it follows that the transfer functions
$H_{n}(z)$ can be expressed as
\begin{equation}
H_{n}(z)=z^{-n}\sum_{k=0}^{N-1}H(k)G_{k}(z)F_{kn}\big(z^{M}\big),\label{eq:Hn}
\end{equation}
where $F_{kn}(z)$ denote the polyphase components of $F_{k}(z)$
in the $M$-fold polyphase decomposition
\begin{equation}
F_{k}(z)=\sum_{n=0}^{M-1}z^{-n}F_{kn}\big(z^{M}\big).\label{eq:Fk_polyphase}
\end{equation}

\begin{table*}
\caption{Properties of the Impulse Responses $h_{n}(q),\,n=0,1,\ldots,M-1$,
for the Overlap-Add and Overlap-Save Methods.}

\begin{centering}
\begin{tabular}{|c|c|c|c|}
\hline 
Property & Effective length & Impulse responses & Impulse responses\tabularnewline
 &  & with quantized $H(k)$ & with quantized $f_{k}(n)$ and/or $g_{k}(n)$\tabularnewline
\hline 
\hline 
Overlap-add, $N\neq KM$ & $M\times\left\lfloor (N-1-n)/M\right\rfloor +M$ & Not circularly shifted & Not circularly shifted\tabularnewline
\hline 
Overlap-add, $N=KM$ & $N$ & Circularly shifted & Not circularly shifted\tabularnewline
\hline 
Overlap-save, all $N$ & $N$ & Circularly shifted & Not circularly shifted\tabularnewline
\hline 
\end{tabular}
\par\end{centering}
\label{Table: Properties}
\end{table*}

Conversely, we can also express the distortion and aliasing functions
of the MFB representation in terms of $H_{n}(z)$. Using again well-known
input-output relations of MFBs \cite{Vaidyanathan_93}, it follows
that $V_{p}(e^{j\omega})$ can be expressed as
\begin{equation}
V_{p}(e^{j\omega})=\frac{1}{M}\sum_{n=0}^{M-1}H_{n}\big(e^{j(\omega-2\pi p/M)}\big)e^{-j2\pi pn/M}.
\end{equation}
It is seen that the distortion frequency response $V_{0}(e^{j\omega})$
is the average of the $M$ filter frequency responses $H_{n}(e^{j\omega})$,
whereas each aliasing frequency response $V_{p}(e^{j\omega})$, $p=1,2,\ldots,M-1,$
is the average of frequency-shifted and rotated (due to the multiplication
by $e^{-j2\pi pn/M}$) versions of $H_{n}(e^{j\omega})$. This means
that a metric based on $H_{n}(e^{j\omega})$ instead of $V_{p}(e^{j\omega})$
may be a better indicator of the worst-case time-domain error of the
overall system. This will be exemplified in Section \ref{sec:Design-Example}.

\section{Impulse Response Properties\label{sec:Impulse-Response-Properties}}

Using finite-wordlength coefficients, i.e., quantized DFT filter coefficients
$H(k)$ and/or quantized complex exponentials in the DFT/IDFT, i.e.,
quantized $g_{k}(n)$ and/or $f_{k}(n)$ (see Footnote 1), the overlap-add
and overlap-save methods have different properties regarding the lengths
of and relations between the $M$ impulse responses $h_{n}(q),\,n=0,1,\ldots,M-1$.
The properties are summarized in Table \ref{Table: Properties}. They
can be deduced from \eqref{eq:Hn} which will be shown and discussed
in detail below in subsections \ref{subsec:Overlap-Add} and \ref{subsec:Overlap-Save}.
For convenience, both filter length and order will be used in those
sections, recalling that the length is the order plus one. Further,
the effective order (and length) will be considered. For an FIR filter
with non-zero impulse response values $h(n)$ for $n=n_{1},n_{1}+1,\ldots,n_{2}$,
the effective order is $n_{2}-n_{1}$, and thus the effective length
is $n_{2}-n_{1}+1$.

\subsection{Overlap-Add\label{subsec:Overlap-Add}}

For the overlap-add method, the length of $g_{k}(n)$ is $M$ whereas
the length of $f_{k}(n)$ is $N$. It is seen in \eqref{eq:Hn} that
$H_{n}(z)$ depends on $G_{k}(z)$ and the polyphase components of
$F_{k}(z)$, i.e., $F_{kn}(z^{M})$ as given by \eqref{eq:Fk_polyphase}.
This means that the effective filter order\footnote{The effective filter order is in general $K_{n}$. However, it can
be smaller which, in particular, occurs when all coefficients are
unquantized. In that case, except for a delay of $M-1$ samples, all
$H_{n}(z)$ coincide with the originally designed filter $H(z)$ whose
order is $L-1$. }, say $K_{n}$, of $H_{n}(z)$ is $K_{n}=M-1+K_{F_{n}}M$ which corresponds
to the order of $H(k)G_{k}(z)F_{kn}(z^{M})$, where $M-1$ is the
order of all $G_{k}(z)$ whereas $K_{F_{n}}$ denote the orders of
$F_{kn}(z)$. The highest-power term in \eqref{eq:Hn} is however
$K_{n}+n$ due to the multiplication of $z^{-n}$. Thus, in the time
domain, each impulse response $h_{n}(q)$ is obtained through an $n$-step
right-shift of the impulse response of $H(k)G_{k}(z)F_{kn}(z^{M})$.
It will thus have $n-1$ initial zero-valued impulse response values.
Furthermore, the effective order $K_{n}$ depends on $n$ in general.
This is because $K_{F_{n}}$ are generally not the same for all $n$.
The exception is when $N$ is an integer multiple of $M$, say $N=KM$,
in which case $K_{F_{n}}=K-1$ for all $n$ and, consequently, $K_{n}=M-1+(K-1)M=N-1$
for all $n$. In general, when $N\neq KM$, $K_{F_{n}}=\left\lfloor (N-1-n)/M\right\rfloor $.

Further, since the order of $G_{k}(z)$ is $M-1$, and $G_{k}(z)$
is multiplied by $F_{kn}(z^{M})$ when expressing the overall transfer
function $H_{n}(z)$ in \eqref{eq:Hn}, the impulse response of $z^{n}H_{n}(z)$,
say $d_{n}(q)$, only depends on $g_{k}(m)$ for one distinct time
index $m$ for each $q$. To be precise, it follows from \eqref{eq:gk_oa},
\eqref{eq:fk_oa}, and \eqref{eq:Hn} that $d_{n}(q)$, $q=0,1,\ldots,K_{n}$,
can be expressed as
\begin{equation}
d_{n}(q)=\sum_{k=0}^{N-1}H(k)\sum_{r=0}^{K_{F_{n}}}g_{k}(q-rM)f_{k}(n+rM).\label{eq:d_n(q)_OA}
\end{equation}
Since the length of $g_{k}(q)$ is $M$, $g_{k}(q-rM)$ correspond
to nonoverlapping right-shifted (by $M$) versions of $g_{k}(q)$.
Hence, for each $q$, only one term in the right-most sum in \eqref{eq:d_n(q)_OA}
is non-zero. 

For the special case when $N=KM$, and thus $K_{F_{n}}=K-1$, $d_{n}(q)$
can be written as
\begin{eqnarray}
d_{n}(q) & = & \sum_{k=0}^{N-1}H(k)\sum_{r=0}^{K-1}g_{k}(q-rM)f_{k}(n+rM)\nonumber \\
 & = & \frac{1}{N}\sum_{k=0}^{N-1}H(k)\nonumber \\
 & \times & \sum_{r=0}^{K-1}e^{j2\pi(q-M+1-rM)k/N}e^{j2\pi(n+rM)k/N}\nonumber \\
 & = & \frac{1}{N}\sum_{k=0}^{N-1}H(k)e^{j2\pi(q-M+1+n)k/N}.\label{eq:d_n(q)_OA2}
\end{eqnarray}
The last equality holds since only one term in the $K$-term summation
is non-zero for each $q$. When $H(k)$ are quantized, but $g_{k}(q)$
and $f_{k}(q)$ are not quantized, $d_{n}(q)$ are circularly shifted
\cite{Mitra_06} versions of each other, which means that all $M$
impulse responses contain the same set of $N$ values. To show this,
consider $d_{n+m}(q)$ which amounts to replacing $n$ with $n+m$
in \eqref{eq:d_n(q)_OA2}. As seen in \eqref{eq:d_n(q)_OA2}, this
is equivalent to replacing $q$ with $q+m$, which corresponds to
$d_{n}(q)$ circularly shifted to the left by $m$ samples. It is
also noted that $-M+1$ on the last two lines of \eqref{eq:d_n(q)_OA2}
emanates from the additional delay of $M-1$ samples due to the block
processing, as mentioned in Footnote 5.

In the general case, when $N\neq KM$, $K_{F_{n}}$are not the same
for all $n$, in which case \eqref{eq:d_n(q)_OA} is still valid,
but not \eqref{eq:d_n(q)_OA2}. Here, since all $d_{n}(q)$ do not
have the same length, the circular-shift property is lost. The property
is also lost for all $N$ when $g_{k}(q)$ and $f_{k}(q)$ are quantized,
meaning that the complex exponentials in \eqref{eq:d_n(q)_OA2} are
quantized (see Footnote 1). This is because the two independently
quantized exponentials, on the second last line in \eqref{eq:d_n(q)_OA2},
will have index-dependent quantization errors and the last equality
and equivalence utilized above will then no longer hold.

\begin{table*}[t]
\caption{Example 1: Original Impulse Response $h(q)$ (Right-Shifted) and Overlap-Add
Impulse Responses $h_{n}(q),\,n=0,1,2,3$, Using Quantized $H(k)$
but Unquantized $g_{k}(n)$ and $f_{k}(n)$, Illustrating That the
Circular-Shift Property Does Not Hold When $N\protect\neq KM$.}

\begin{centering}
\begin{tabular}{|c|c|c|c|c|}
\hline 
Original, $h(q-3)$ & $h_{0}(q)$ & $h_{1}(q)$ & $h_{2}(q)$ & $h_{3}(q)$\tabularnewline
\hline 
\hline 
0 & 0.000815299395028 & 0 & 0 & 0\tabularnewline
\hline 
0 & 0.000030422174521 & 0.000030422174521 & 0 & 0\tabularnewline
\hline 
0 & 0.000083095006610 & 0.000083095006610 & 0.000083095006610 & 0\tabularnewline
\hline 
-0.065517977199101 & -0.064843750000000 & -0.064843750000000 & -0.064843750000000 & -0.064843750000000\tabularnewline
\hline 
0.054777425047761 & 0.054418477371339 & 0.054418477371339 & 0.054418477371339 & 0.054418477371339\tabularnewline
\hline 
0.314937451772624 & 0.314709622812781 & 0.314709622812781 & 0.314709622812781 & 0.314709622812781\tabularnewline
\hline 
0.464142316077418 & 0.464214378227023 & 0.464214378227023 & 0.464214378227023 & 0.464214378227023\tabularnewline
\hline 
0.314937451772624 & 0.315733563910444 & 0.315733563910444 & 0.315733563910444 & 0.315733563910444\tabularnewline
\hline 
0.054777425047761 & 0.054687500000000 & 0.054687500000000 & 0.054687500000000 & 0.054687500000000\tabularnewline
\hline 
-0.065517977199101 & -0.065629858897746 & -0.065629858897746 & -0.065629858897746 & -0.065629858897746\tabularnewline
\hline 
0 & 0.000815299395028 & 0.000815299395028 & 0 & 0.000815299395028\tabularnewline
\hline 
0 & 0.000030422174521 & 0.000030422174521 & 0 & 0\tabularnewline
\hline 
0 & 0 & 0.000083095006611 & 0 & 0\tabularnewline
\hline 
\end{tabular}
\par\end{centering}
\label{Table:Ex1_OA}
\end{table*}

\begin{table*}
\caption{Example 1: Original Impulse Response $h(q)$ (Right-Shifted) and Overlap-Save
Impulse Responses $h_{n}(q),\,n=0,1,2,3$, Using Quantized $H(k)$
but Unquantized $g_{k}(n)$ and $f_{k}(n)$, Showing the Circular-Shift
Property.}

\begin{centering}
\begin{tabular}{|c|c|c|c|c|}
\hline 
Original, $h(q-3)$ & $h_{0}(q)$ & $h_{1}(q)$ & $h_{2}(q)$ & $h_{3}(q)$\tabularnewline
\hline 
\hline 
0 & 0.000815299395028 & 0 & 0 & 0\tabularnewline
\hline 
0 & 0.000030422174521 & 0.000030422174521 & 0 & 0\tabularnewline
\hline 
0 & 0.000083095006610 & 0.000083095006610 & 0.000083095006610 & 0\tabularnewline
\hline 
-0.065517977199101 & -0.064843750000000 & -0.064843750000000 & -0.064843750000000 & -0.064843750000000\tabularnewline
\hline 
0.054777425047761 & 0.054418477371339 & 0.054418477371339 & 0.054418477371339 & 0.054418477371339\tabularnewline
\hline 
0.314937451772624 & 0.314709622812781 & 0.314709622812781 & 0.314709622812781 & 0.314709622812781\tabularnewline
\hline 
0.464142316077418 & 0.464214378227023 & 0.464214378227023 & 0.464214378227023 & 0.464214378227023\tabularnewline
\hline 
0.314937451772624 & 0.315733563910444 & 0.315733563910444 & 0.315733563910444 & 0.315733563910444\tabularnewline
\hline 
0.054777425047761 & 0.054687500000000 & 0.054687500000000 & 0.054687500000000 & 0.054687500000000\tabularnewline
\hline 
-0.065517977199101 & -0.065629858897746 & -0.065629858897746 & -0.065629858897746 & -0.065629858897746\tabularnewline
\hline 
0 & 0 & 0.000815299395028 & 0.000815299395028 & 0.000815299395028\tabularnewline
\hline 
0 & 0 & 0 & 0.000030422174521 & 0.000030422174521\tabularnewline
\hline 
0 & 0 & 0 & 0 & 0.000083095006610\tabularnewline
\hline 
\end{tabular}
\par\end{centering}
\label{Table:Ex1_OS_a}
\end{table*}

\begin{table*}
\caption{Example 1: Original Impulse Response $h(q)$ (Right-Shifted) and Overlap-Save
Impulse Responses $h_{n}(q),\,n=0,1,2,3$, Using Quantized $H(k)$,
$g_{k}(n)$, and $f_{k}(n)$, Showing That the Circular-Shift Property
Is Lost.}

\begin{centering}
\begin{tabular}{|c|c|c|c|c|}
\hline 
Original, $h(q-3)$ & $h_{0}(q)$ & $h_{1}(q)$ & $h_{2}(q)$ & $h_{3}(q)$\tabularnewline
\hline 
\hline 
0 & 0.001343357563019 & 0 & 0 & 0\tabularnewline
\hline 
0 & 0.000361371040344 & 0.000361371040344 & 0 & 0\tabularnewline
\hline 
0 & 0.000538158416748 & 0.000160551071167 & 0.000538158416748 & 0\tabularnewline
\hline 
-0.065517977199101 & -0.064286172389984 & -0.064312195777893 & -0.064312195777893 & -0.064286172389984\tabularnewline
\hline 
0.054777425047761 & 0.054309082031250 & 0.054947161674500 & 0.054378080368042 & 0.054947161674499\tabularnewline
\hline 
0.314937451772624 & 0.313703811168671 & 0.314443969726562 & 0.314058876037598 & 0.314058876037598\tabularnewline
\hline 
0.464142316077418 & 0.463059282302857 & 0.463059282302857 & 0.463626098632813 & 0.463081991672516\tabularnewline
\hline 
0.314937451772624 & 0.315321087837219 & 0.314716339111328 & 0.315321087837219 & 0.315228271484375\tabularnewline
\hline 
0.054777425047761 & 0.054968869686127 & 0.054803586006165 & 0.054803586006165 & 0.054968869686127\tabularnewline
\hline 
-0.065517977199101 & -0.065100097656250 & -0.065209484100342 & -0.065339374542236 & -0.065209484100342\tabularnewline
\hline 
0 & 0 & 0.001248168945313 & 0.000872826576233 & 0.000872826576233\tabularnewline
\hline 
0 & 0 & 0 & 0.000271606445313 & 0.000208508968353\tabularnewline
\hline 
0 & 0 & 0 & 0 & 0.000347900390625\tabularnewline
\hline 
\end{tabular}
\par\end{centering}
\label{Table:Ex1_OS_b}
\end{table*}

\subsection{Overlap-Save\label{subsec:Overlap-Save}}

Here, the order of $F_{k}(z)$ is $M-1$, which means that the order
of all polyphase components $F_{kn}(z^{M})$ is zero, whereas the
order of $G_{k}(z)$ is $N-1$. Consequently, the effective order
is $K_{n}=N-1$ for all $n$. This coincides with the special case
of the overlap-add method with $N=KM$.

Further, when $H(k)$ are quantized, but $g_{k}(q)$ and $f_{k}(q)$
are unquantized, the impulse responses of $z^{n}H_{n}(z)$, $d_{n}(q)$,
are circularly shifted versions of each other regardless of the values
of $N$ and $M$. This is different from the overlap-add method for
which this property holds only when $N=KM$. For the overlap-save
method, it holds for all $N$ and $M$ because the order of all polyphase
components $F_{kn}(z^{M})$ is zero. Consequently, each $F_{kn}(z^{M})$
is here a constant, viz. $F_{kn}(z^{M})=f_{k}(n)$, and it then follows
from \eqref{eq:gk_os}, \eqref{eq:fk_os}, and \eqref{eq:Hn}, that
$d_{n}(q)$ can be written as
\begin{eqnarray}
d_{n}(q) & = & \sum_{k=0}^{N-1}H(k)g_{k}(q)f_{k}(n)\nonumber \\
 & = & \frac{1}{N}\sum_{k=0}^{N-1}H(k)e^{j2\pi(q+1)k/N}e^{j2\pi(n-M)k/N}\nonumber \\
 & = & \frac{1}{N}\sum_{k=0}^{N-1}H(k)e^{j2\pi(q+1+n-M)k/N}.\label{eq:d_n(q)_OS}
\end{eqnarray}
Consider now $d_{n+m}(q)$ which amounts to replacing $n$ with $n+m$
in \eqref{eq:d_n(q)_OS}. As seen in \eqref{eq:d_n(q)_OS}, this is
equivalent to replacing $q$ with $q+m$, which corresponds to $d_{n}(q)$
circularly shifted to the left by $m$ samples. As for the overlap-add
method, the property is however lost when $g_{k}(q)$ and $f_{k}(q)$
are quantized. 

\section{Design Examples\label{sec:Design-Example}}

\textit{Example 1.} This example will illustrate the impulse response
properties provided in Section \ref{sec:Impulse-Response-Properties}.
To this end, we use a linear-phase FIR filter of length $L=7$, a
block length of $M=4$ and a DFT length of $N=10$. We have used an
equiripple design, assuming passband and stopband edges at $0.3\pi$
and $0.6\pi$, respectively, and equal passband and stopband ripples.
When rounding, we have used $8$ fractional bits. 

Table \ref{Table:Ex1_OA} gives the original impulse response $h(q)$
(right-shifted three steps to ease the comparison) and the four impulse
responses $h_{n}(q),\,n=0,1,2,3$, when using the overlap-add method
and with $H(k)$ quantized. It is seen that the impulse responses
have different effective lengths and that the circular-shift property
does not hold which is because $N\neq KM$. Table \ref{Table:Ex1_OS_a}
gives the corresponding impulse responses for the overlap-save method.
Here, it is seen that the impulse responses have the same effective
length and that the circular-shift property holds. However, as seen
in Table \ref{Table:Ex1_OS_b}, when $g_{k}(n)$ and $f_{k}(n)$ are
quantized as well (i.e., the complex exponentials in the DFT/IDFT
are quantized, see Footnote 1), also the overlap-save method loses
the circular-shift property, but the impulse responses are still of
the same effective length.

\begin{figure}[t!]
\centering \scalebox{0.8}{\includegraphics[scale=0.75]{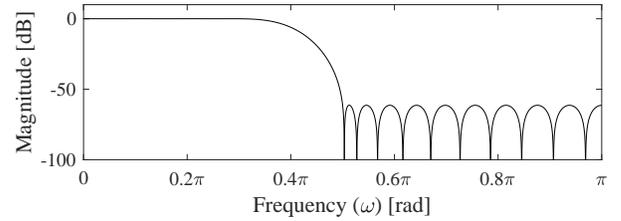}}
\caption{Examples 2: Initial infinite-precision filter frequency response $H(e^{j\omega})$.}

\label{fig:Ex1_H} 
\end{figure}

\begin{figure}[t!]
\centering \scalebox{0.8}{\includegraphics[scale=0.75]{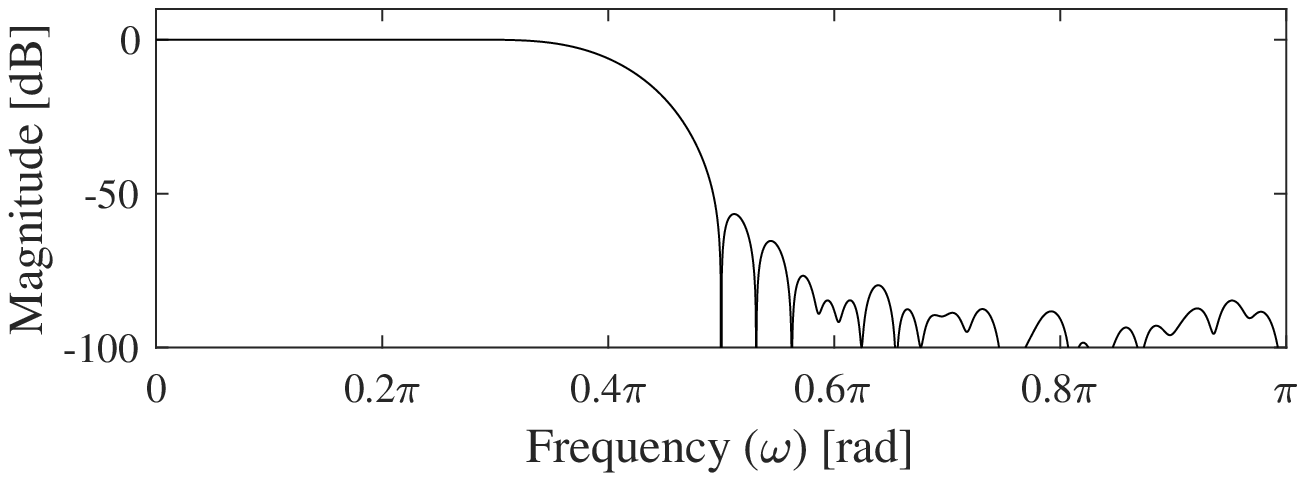}}
\caption{Example 2: Distortion frequency response $V_{0}(e^{j\omega})$. }

\label{fig:Ex1_H_average} 
\end{figure}

\begin{figure}[t!]
\centering \scalebox{0.8}{\includegraphics[scale=0.75]{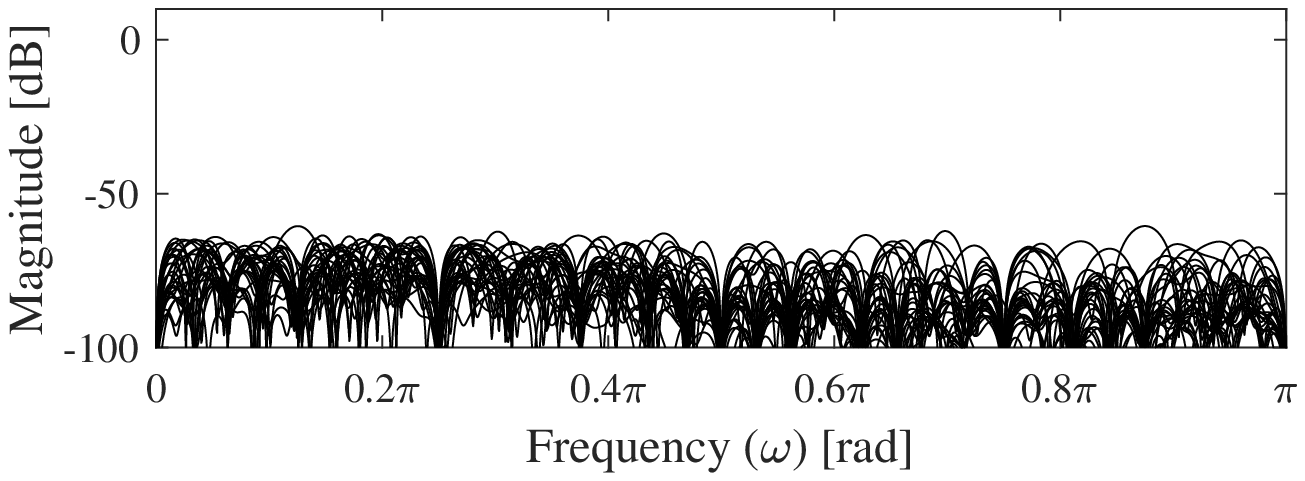}}
\caption{Example 2: Aliasing frequency responses $V_{p}(e^{j\omega})$, $p=1,2,\ldots,M-1$. }

\label{fig:Ex1_Aliasing} 
\end{figure}

\begin{figure}[t!]
\centering \scalebox{0.8}{\includegraphics[scale=0.75]{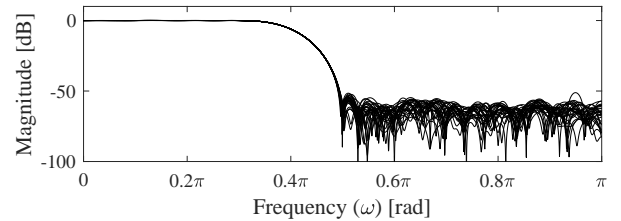}}
\caption{Example 2: Frequency responses $H_{n}(e^{j\omega})$, $n=0,1,\ldots,M-1$.}

\label{fig:Ex1_H_all} 
\end{figure}

\textit{Example 2:} This example will illustrate the frequency-domain
properties of the MFB and PTVIR representations. To this end, we first
design an equiripple linear-phase FIR filter of length $L=35$ and
with passband and stopband edges at $0.3\pi$ and $0.5\pi$, respectively,
and passband and stopband ripples of $0.001$ ($-60$ dB). The frequency
response $H(e^{j\omega})$ of the initial filter with infinite precision
(here Matlab precision) impulse response values (coefficients) $h$($n$)
is seen in Fig. \ref{fig:Ex1_H}. Next, we implement the filter with
the overlap-add method (Fig. \ref{fig:OA-DFT}) with $M=30$, and
thus $N=64$, and with eight fractional bits for $H(k)$ as well as
for $g_{k}(n)$ and $f_{k}(n)$. The resulting distortion and aliasing
frequency responses $V_{0}(e^{j\omega})$ and $V_{p}(e^{j\omega})$,
$p=1,2,\ldots,M-1$, in the MFB representation (Fig. \ref{fig:FB_representation})
are seen in Figs. \ref{fig:Ex1_H_average} and \ref{fig:Ex1_Aliasing},
respectively. The corresponding frequency responses $H_{n}(e^{j\omega})$
in the PTVIR representation (Fig. \ref{fig:TV_representation}) are
plotted in Fig. \ref{fig:Ex1_H_all}. As seen, the worst-case responses
of $H_{n}(e^{j\omega})$ are some $10$ dB larger than the aliasing
terms in the stopband. This illustrates that, in applications where
the worst-case time-domain error (difference between the actual output
$y(n)$ and the desired one) is more important than the average error,
a metric based on $H_{n}(e^{j\omega})$ instead of $V_{p}(e^{j\omega})$
is more appropriate.

\begin{figure}[t!]
\centering \scalebox{0.8}{\includegraphics[scale=0.75]{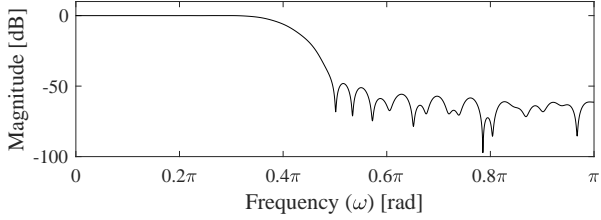}}
\caption{Example 2: Distortion frequency response $V_{0}(e^{j\omega})$ when
using a reduced DFT length. }

\label{fig:Ex2_H_avarage} 
\end{figure}

\begin{figure}[t!]
\centering \scalebox{0.8}{\includegraphics[scale=0.75]{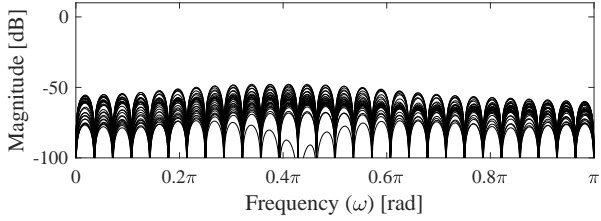}}
\caption{Example 2: Aliasing frequency responses $V_{p}(e^{j\omega})$, $p=1,2,\ldots,M-1$,
when using a reduced DFT length. }

\label{fig:Ex2_Aliasing} 
\end{figure}

\begin{figure}[t!]
\centering \scalebox{0.8}{\includegraphics[scale=0.75]{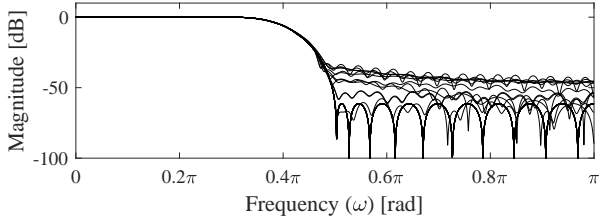}}
\caption{Example 2: Frequency responses $H_{n}(e^{j\omega})$, $n=0,1,\ldots,M-1$,
when using a reduced DFT length.}

\label{fig:Ex2_H_all} 
\end{figure}

To further illustrate the difference between the MFB and PTVIR representations,
we perform the same analysis as above, but here with a reduced DFT
length of $N=56$ instead of quantized coefficients. The corresponding
frequency responses are plotted in Figs. \ref{fig:Ex2_H_avarage}--\ref{fig:Ex2_H_all}.
It is seen that the difference between the two representations is
more pronounced in this case as the difference between the best and
worst $H_{n}(e^{j\omega})$ is quite large. It also illustrates that
one can only use slightly shorter DFT lengths than theoretically required.
Otherwise, the performance degradation becomes very large.

\textit{Example 3: }Periodic signals with frequencies matching the
frequencies of a DFT of length $N/P$, can be efficiently time-domain
interpolated through the use of a DFT and IDFT together with zero
padding in the frequency domain. The basic principle is to, blockwise,
compute a length-($N/P$) DFT of the input signal and then use the
so obtained DFT coefficients as $N/P$ appropriately allocated nonzero-valued
DFT coefficients, together with $N-N/P$ zero-valued DFT coefficients,
in a length-$N$ DFT. Finally, a length-$N$ IDFT is computed, generating
$N$ time-domain sample values, which corresponds to the original
signal interpolated by $P$. Without quantized coefficients, the interpolation
is error free for these periodic signals. However, when the signal
is not periodic within a block of $N/P$ ($N$) samples before (after)
the interpolation, large errors are introduced. 

\begin{figure}[t!]
\centering \scalebox{0.8}{\includegraphics[scale=0.75]{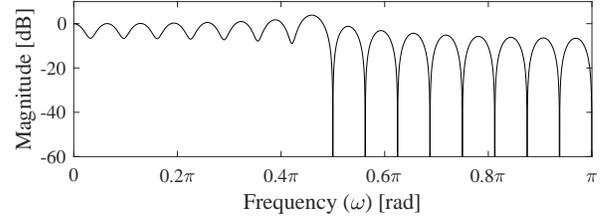}}
\caption{Example 3: Filter frequency response $H(e^{j\omega})$.}

\label{fig:Ex3_H}
\end{figure}

\begin{figure}[t!]
\centering \scalebox{0.8}{\includegraphics[scale=0.75]{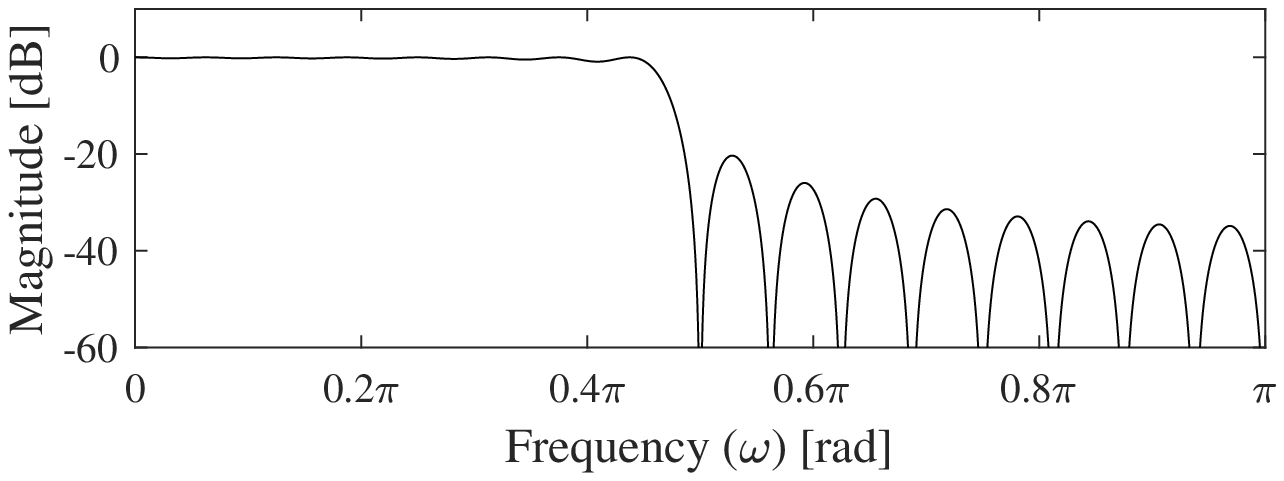}}
\caption{Example 3: Distortion frequency response $V_{0}(e^{j\omega})$.}

\label{fig:Ex3_Distortion}
\end{figure}

\begin{figure}[t!]
\centering \scalebox{0.8}{\includegraphics[scale=0.75]{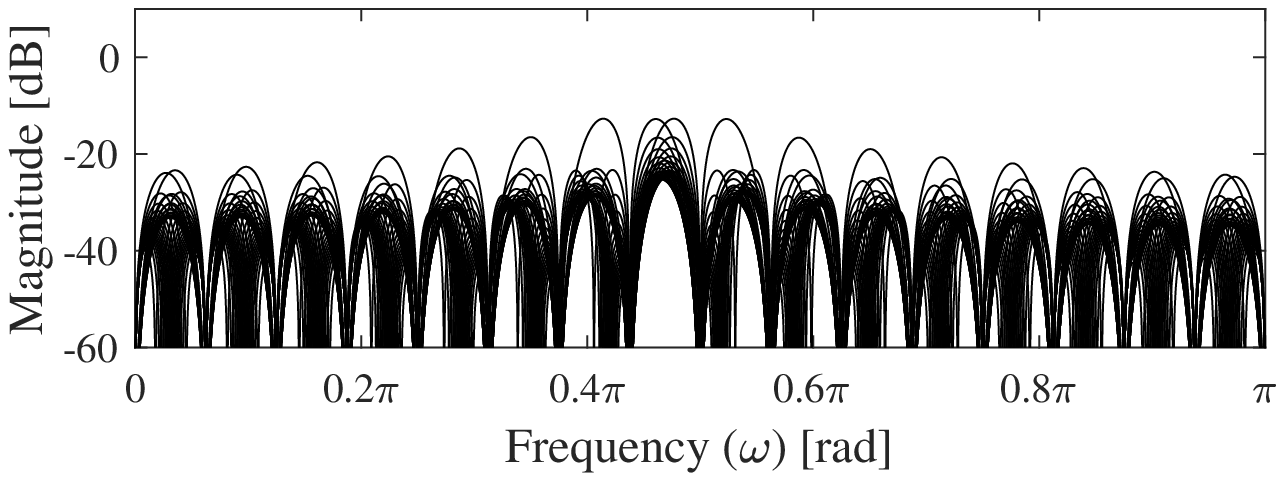}}
\caption{Example 3: Aliasing frequency responses $V_{p}(e^{j\omega})$, $p=1,2,\ldots,M-1$.}

\label{fig:Ex3_Aliasing}
\end{figure}

To illustrate the interpolation error for nonperiodic signals, it
is first recognized that the scheme explained above is equivalent
to first upsampling the signal by $P$, and then use the upsampled
signal as the input to the overlap-add or overlap-save implementation
with $N=L=M$ and with $H(k)=1$ ($H(k)=0$) for $k$-values corresponding
to the nonzero-valued (zero-valued) DFT coefficients. For interpolated
signals with frequencies between $2\pi k/N$, $k=0,1,\ldots,N-1$,
large interpolation errors are introduced for two reasons. Firstly,
the frequency response of the underlying filter, with the impulse
response $h(n)$ obtained from the IDFT of $H(k)$, is poor between
the frequencies $2\pi k/N$. This is illustrated in Fig. \ref{fig:Ex3_H}
for the case where $P=2$ and $N=32$. Secondly, since $N=L=M$, the
length of the DFT, $N$, is shorter than required ($L+M-1$) for a
proper implementation of linear convolution. This is seen in Figs.
\ref{fig:Ex3_Distortion} and \ref{fig:Ex3_Aliasing} which plot the
distortion and aliasing functions, respectively. In a proper implementation
($N=L+M-1$) with unquantized coefficients, the aliasing functions
are zero and the distortion function equals the frequency response
of the underlying length-$L$ filter impulse response for all frequencies,
not only for the frequencies $2\pi k/N$.

\begin{figure}[t!]
\centering \scalebox{0.8}{\includegraphics[scale=0.75]{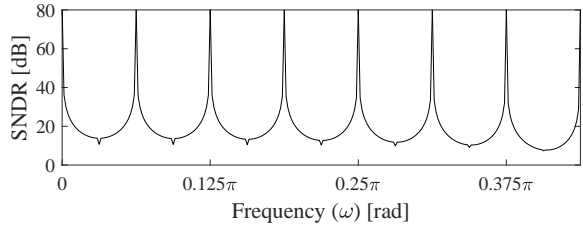}}
\caption{Example 3: SNDR as a function of the frequency of the interpolated
signal.}

\label{fig:Ex3_SNDR}
\end{figure}

\begin{figure}[t!]
\centering \scalebox{0.8}{\includegraphics[scale=0.75]{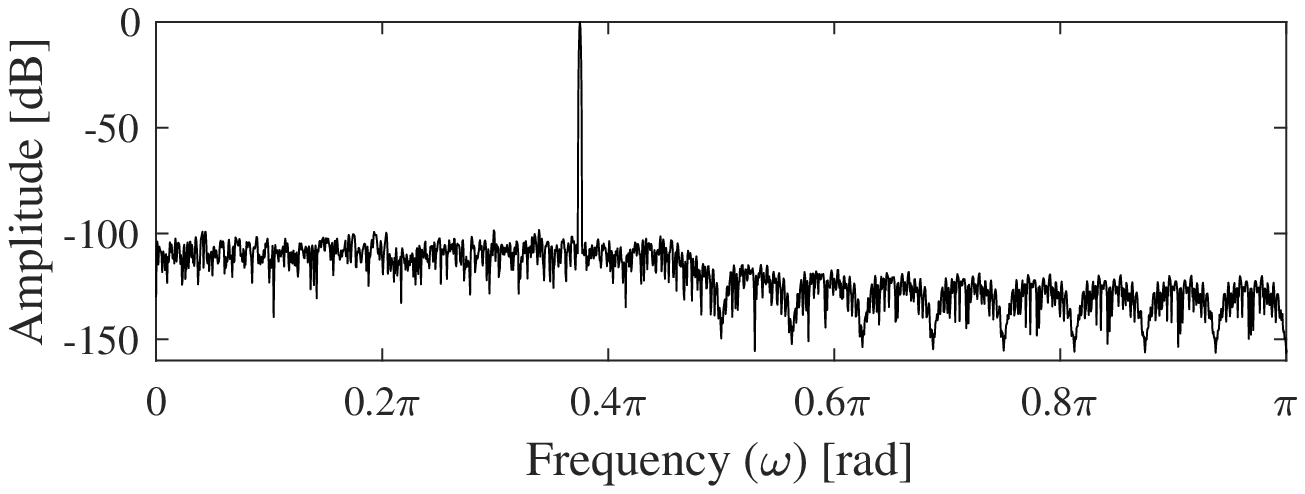}}
\caption{Example 3: Spectrum of the interpolated signal when its frequency
is $2\pi\times6/32$.}

\label{fig:Ex3_spectrum_1}
\end{figure}

\begin{figure}[t!]
\centering \scalebox{0.8}{\includegraphics[scale=0.75]{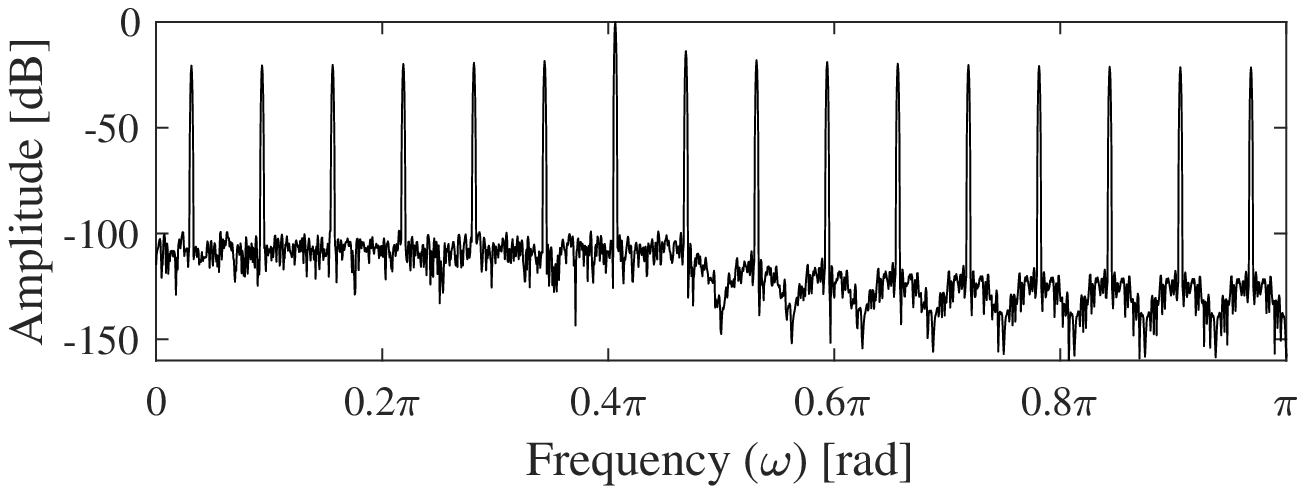}}
\caption{Example 3: Spectrum of the interpolated signal when its frequency
is $2\pi\times6.5/32$.}

\label{fig:Ex3_spectrum_2}
\end{figure}

The two sources of errors for nonperiodic signals result in large
interpolation errors. This is illustrated in Fig. \ref{fig:Ex3_SNDR}
which plots the signal-to-noise-and distortion ratio (SNDR) as a function
of frequency, when the input signal is a noisy sinusoid with a signal-to-noise
ratio (SNR) of 80 dB. It is seen that for the frequencies $2\pi k/N$
(periodic signals), the SNDR is 80 dB as the interpolation is then
error free and the SNDR determined by the SNR of the input signal.
For frequencies between $2\pi k/N$ (nonperiodic signals), the SNDR
is poor, especially around the mid-point between adjacent values of
$2\pi k/N$ where it is only some $7$--$14$ dB. Figures \ref{fig:Ex3_spectrum_1}
and \ref{fig:Ex3_spectrum_2} plot the spectrum for two of these signals,
for the frequencies $2\pi\times6/32$ and $2\pi\times6.5/32$. As
the plots show, the desired signal is obtained in the former of these
two cases, whereas large aliasing terms are present in the latter,
located at the signal frequency plus/minus multiples of $2\pi/N$.
These errors match the large aliasing functions seen in Fig. \ref{fig:Ex3_Aliasing}.
In order to use frequency-domain implementation of time-domain interpolation
over the whole frequency range, it is thus necessary to properly design
an interpolation filter and then implement the overlap-add or overlap-save
method properly as in, e.g., \cite{Muramatsu_1997}.

\begin{table*}
\caption{Multiplication Rates. Complex (Real) Means That Both the Signal and
Impulse Response Are Complex-Valued (Real-Valued). Symmetric Means
That the Impulse Response Is Symmetric (Not Conjugate Symmetric).}

\begin{centering}
\begin{tabular}{|c|c|c|c|c|}
\hline 
Case & Complex & Complex symmetric & Real & Real symmetric\tabularnewline
\hline 
\hline 
Time-domain multiplication rate $R_{\textnormal{TD}}$ & $3L$ & $3\left\lceil L/2\right\rceil $ & $L$ & $\left\lceil L/2\right\rceil $\tabularnewline
\hline 
Frequency-domain multiplication rate $R_{\textnormal{FD}}$ & $\frac{2(N\log_{2}(N)-3N/2+4)}{N-L+1}$ & $\frac{2(N\log_{2}(N)-3N/2+4)}{N-L+1}$ & $\frac{N\log_{2}(N)-3N/2+4}{N-L+1}$ & $\frac{N\log_{2}(N)-3N/2+4}{N-L+1}$\tabularnewline
\hline 
\end{tabular}
\par\end{centering}
\label{Table: Complexities}
\end{table*}

\section{Implementation Complexity\label{sec:Implementation-Complexity}}

In this section, we will analyze and compare the computational complexities
of the frequency-domain implementations and the corresponding time-domain
implementations, assuming direct-form FIR filter structures \cite{Jackson_96,Wanhammar_11}
for the latter. As a measure of computational complexity, we use the
multiplication rate which is defined as the number of multiplications
required to compute each output sample. The focus is here on multiplications
as they are generally substantially more costly to implement than
additions.%

Earlier publications on the frequency-domain implementations indicate
that they become more efficient than the corresponding time-domain
implementations for filter lengths greater than $25$--$80$ for
general FIR filters\footnote{The references \cite{Oppenheim_89} and \cite{Lyons_96} indicate
filter lengths $25$--$30$ and $40$--$80$, respectively.}, and thus around $50$--$160$ for linear-phase FIR filters due
to their impulse-response symmetries. However, as will be shown in
this section, the frequency-domain implementations become more efficient
for filter lengths far below those numbers. In part, this is because
the use of more efficient FFT algorithms (in particular split-radix
algorithms) can further reduce the complexity required to implement
the DFT and IDFT. These further savings have been reported in other
publications, e.g., in the context of chromatic-dispersion equalization
\cite{Ishihara_2011} and sampling rate conversion \cite{Muramatsu_1997}.
However, here it will be shown that even further complexity savings
are feasible using optimal DFT lengths which have not been used in
earlier publications. A common selection has been a DFT length that
is twice the filter length \cite{Ishihara_2011}. As will be seen
later in this section, the optimal DFT length is around three times
the filter length for short filters and it increases with the filter
length. In particular, with optimal DFT lengths for general filters
(without symmetries), we will show that the frequency-domain implementations
are more efficient for all filter lengths. This was not seen in \cite{Muramatsu_1997,Ishihara_2011}
where short-length filters were reported to be more efficiently implemented
in the time domain. It is noted though that \cite{Muramatsu_1997}
considers sampling rate conversions (by two in the examples), in which
case the complexity expressions and analysis differ somewhat from
the ones presented here.

\subsection{Complexity Comparison}

Table \ref{Table: Complexities} gives the multiplication rate as
a function of $N$ and $L$ for the frequency-domain and time-domain
implementations, both for complex-valued and real-valued signals and
impulse responses, and for general and symmetric impulse responses.
For the complexity of the FFT and IFFT, we assume that each complex
multiplication is implemented using three real multiplications. Assuming
further that $N=2^{P}$, $P$ integer, and using split-radix algorithms,
each of the FFT and IFFT can then be implemented with $N\log_{2}(N)-3N+4$
real multiplications for a complex-valued signal and impulse response
\cite{Sorensen,Burrus_2018}. For a real-valued signal and impulse
response, the number is halved \cite{Sorensen,Burrus_2018}. Further,
the coefficients $H(k)$ require $3N$ multiplications in the complex
case, but only $3N/2$ in the real case because the outputs of the
FFT as well as $H(k)$ are then conjugate symmetric. Thus, for a real-valued
signal and impulse response, the multiplication rate, say $R_{\textnormal{FD}}$,
becomes 
\begin{equation}
R_{\textnormal{FD}}=\frac{N\log_{2}(N)-3N/2+4}{N-L+1}.\label{eq:M_FD}
\end{equation}
For a complex-valued signal and impulse response, the multiplication
rate is twice the right-hand side in \eqref{eq:M_FD}.

\begin{figure}[t!]
\centering \scalebox{0.8}{\includegraphics[scale=0.75]{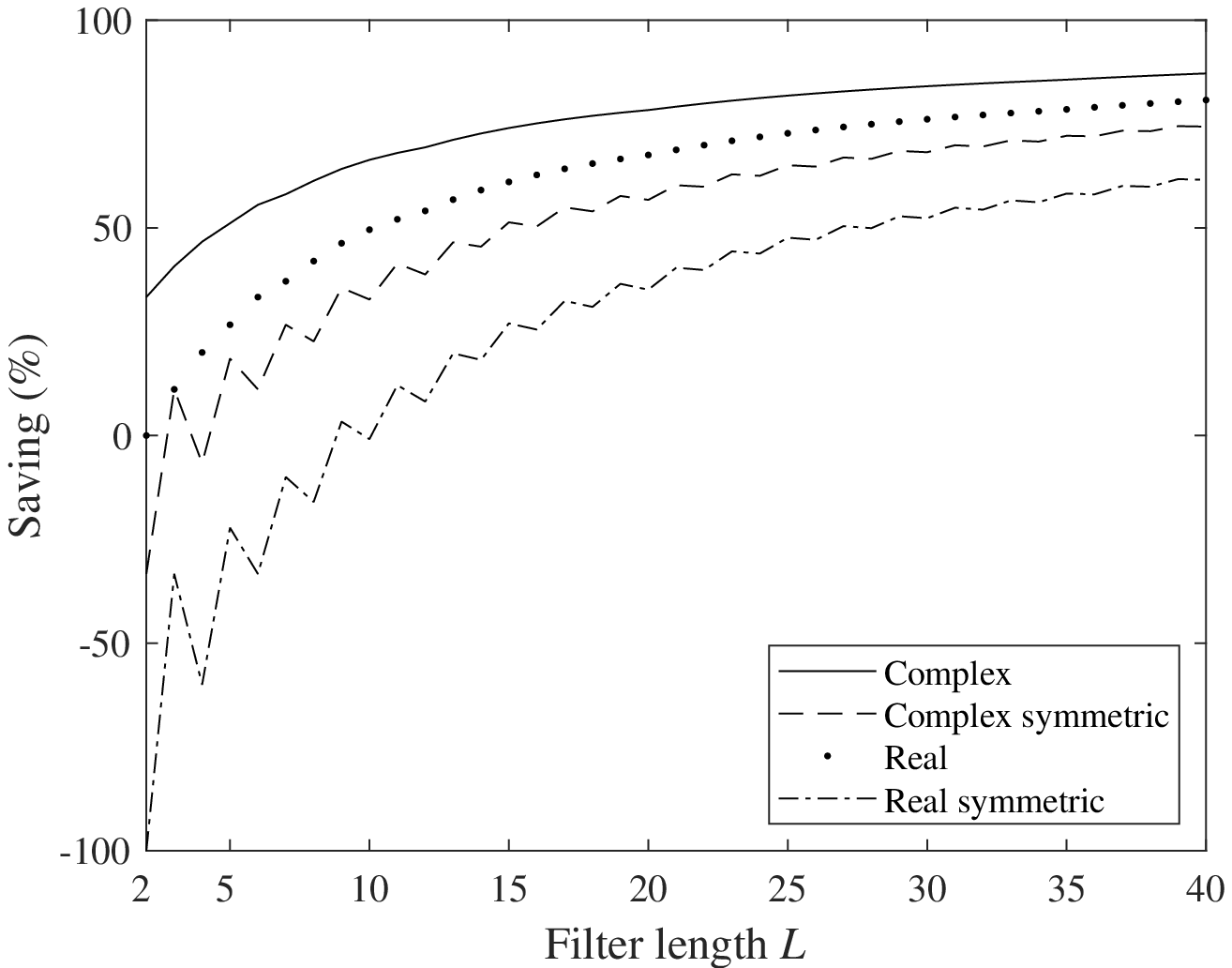}}
\scalebox{0.8}{\includegraphics[scale=0.75]{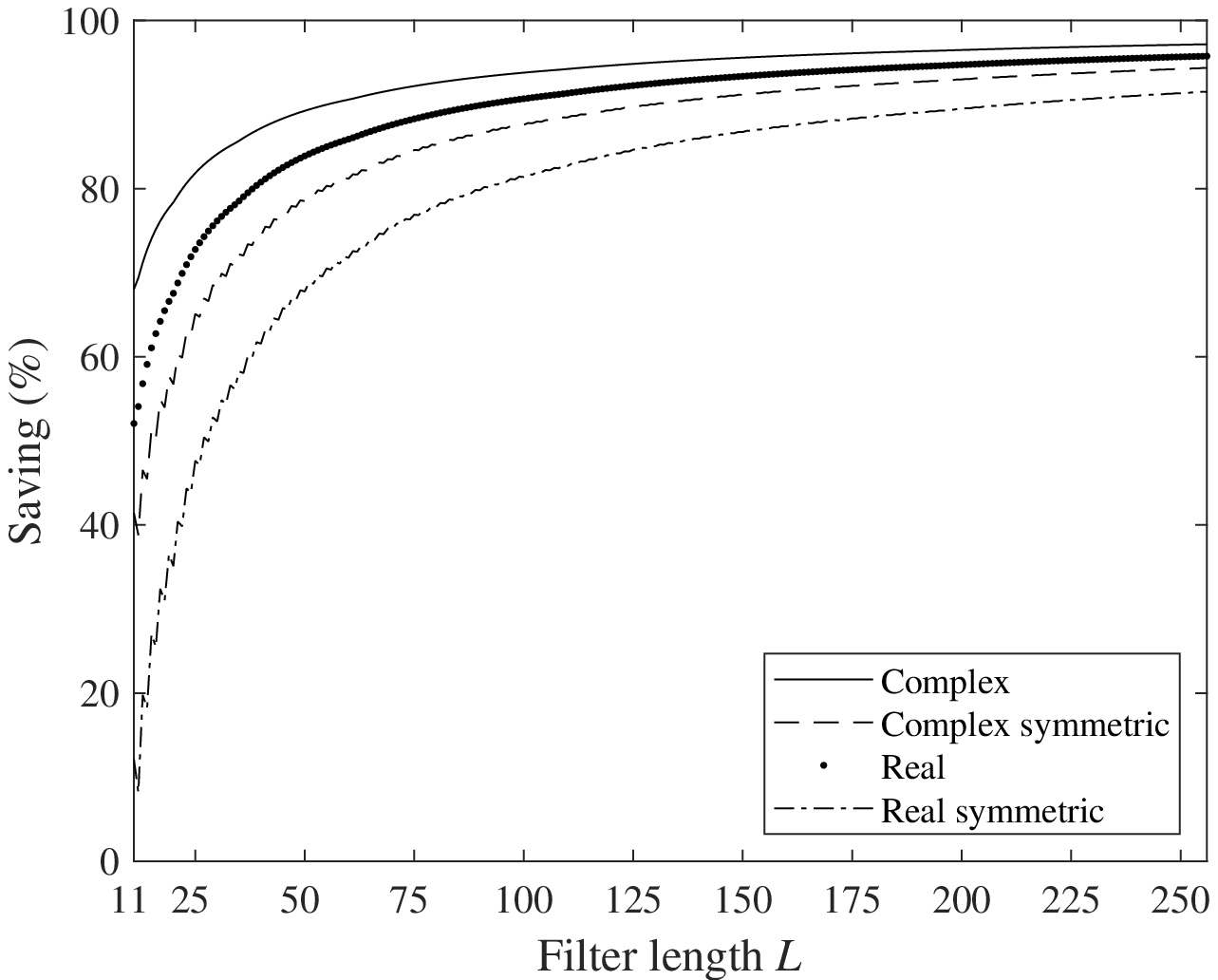}} \caption{Example 4: Computational complexity savings using frequency-domain
implementations instead of time-domain implementations. (It is divided
into two plots for visualization reasons, and there is thus an overlap
for $11\protect\leq L\protect\leq40$).}

\label{fig:Ex4_1}
\end{figure}

Based on the expressions given in Table \ref{Table: Complexities},
Fig. \ref{fig:Ex4_1} plots the savings when using the frequency-domain
implementations instead of the time-domain implementations for $L\in[2,256]$
(divided into two plots for visualization reasons). The saving in
percent is given by $100\times(1-R_{\textnormal{FD}}/R_{\textnormal{TD}})$,
where $R_{\textnormal{TD}}$ denotes the time-domain computational
complexity. Further, for each value of $L$, the optimal saving has
been obtained by minimizing $R_{\textnormal{FD}}$ over different
$N=2^{P}\geq L$ and with $M=N-L+1$. Figure \ref{fig:Ex4_1} shows
that, for the general (unsymmetric) filters, the frequency-domain
implementation is actually superior for all filter lengths. For symmetric
filters, the frequency-domain implementations are computationally
more efficient for filter lengths of $11$ and above in the real case,
and more efficient for odd (even) filter lengths of $3$ ($6$) and
above in the complex case.

\subsection{Estimates of the Complexity}

\begin{figure}[t!]
\centering \scalebox{0.8}{\includegraphics[scale=0.75]{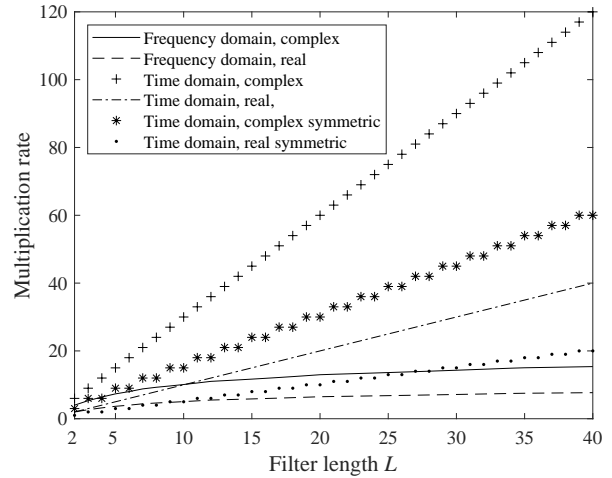}}
\caption{Example 4: Computational complexities using frequency-domain and time-domain
implementations.}

\label{fig:Ex4_1a}
\end{figure}

\begin{figure}[t!]
\centering \scalebox{0.8}{\includegraphics[scale=0.75]{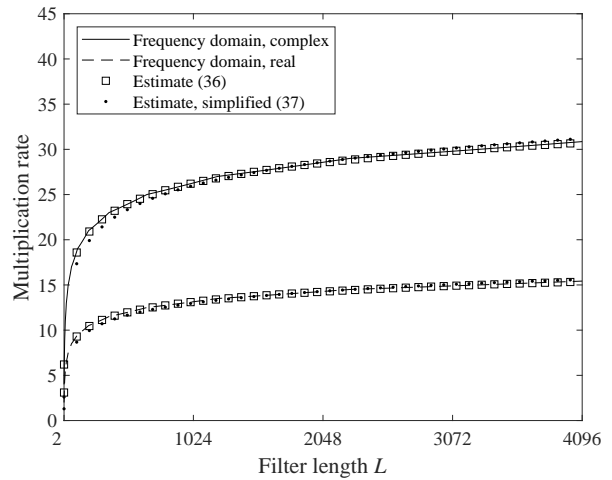}}
\caption{Example 4: Computational complexities using frequency-domain implementations.}

\label{fig:Ex4_1b}
\end{figure}

Figure \ref{fig:Ex4_1a} plots the complexities of the frequency-domain
and time-domain implementations, corresponding to the upper plot in
Fig. \ref{fig:Ex4_1} (i.e., for $L\in[2,40]$). As can be seen, the
computational complexities of the time-domain implementations grow
linearly with $L$, in accordance with the expressions in Table \ref{Table: Complexities}.
For the frequency-domain implementation, the computational complexities
are instead approximately proportional to $\log_{2}(L)$. A good estimation
of the complexity for the real case is
\begin{equation}
\widehat{R}_{\textnormal{FD}}=\frac{\log_{2}(L)+\log_{2}(\log_{2}(L))-\frac{3}{2}+\frac{40}{9(L\times\log_{2}(L))}}{1-\frac{1}{\log_{2}(L)}+\frac{10}{9(L\times\log_{2}(L))}}.\label{eq:M_FD_estim}
\end{equation}
This has been derived by inserting $N=0.9L\log_{2}(L)$ into \eqref{eq:M_FD}
(see the motivation in the last paragraph of this section). Also recall
that the computational complexity is twice as large in the complex
case. Figure \ref{fig:Ex4_1b} plots the computational complexities
of the frequency-domain implementations for $L\in[2,2^{12}]$ ($2^{12}=4096$)
and the corresponding estimations based on \eqref{eq:M_FD_estim}.
It is seen that the estimations are accurate for all values of $L$. 

From \eqref{eq:M_FD_estim}, one can deduce the simplified estimation
\begin{equation}
\widehat{R}_{\textnormal{FD}}=1.3\times\log_{2}(L),\label{eq:M_FD_estim_simpl}
\end{equation}
which is also included in Fig. \ref{fig:Ex4_1b}. It is seen that
it is somewhat less accurate than the expression in \eqref{eq:M_FD_estim},
but it still gives a good approximation of the computational complexity
and it shows that it is approximately proportional to $\log_{2}(L)$.
This also explains the trend of the savings seen in Fig. \ref{fig:Ex4_1}
since the ratio $R_{\textnormal{FD}}/R_{\textnormal{TD}}$ is proportional
to $\log_{2}(L)/L$ which approaches zero when $L$ increases. Thus,
the savings approach one when $L$ increases.

\begin{figure}[t!]
\centering \scalebox{0.8}{\includegraphics[scale=0.75]{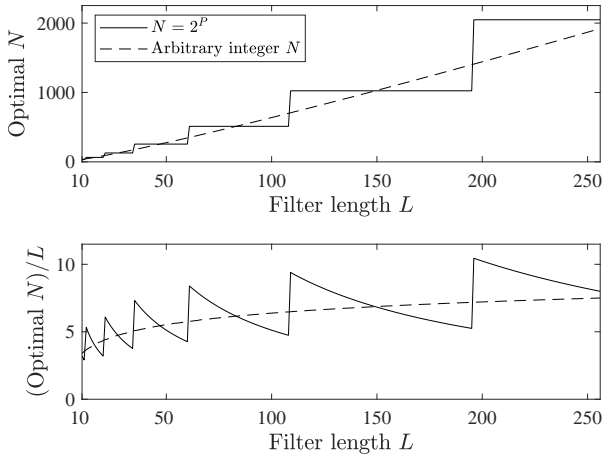}}
\caption{Example 4: DFT length $N$ versus filter length $L$.}

\label{fig:Ex4_3}
\end{figure}

\begin{figure}[t!]
\centering \scalebox{0.8}{\includegraphics[scale=0.75]{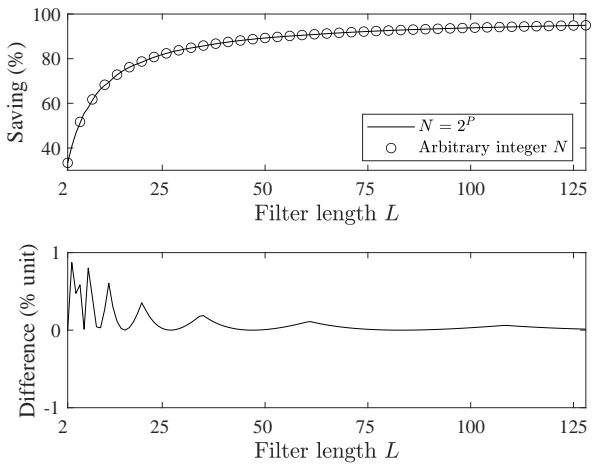}}
\caption{Example 4: Computational complexity savings for $N=2^{P}$ and arbitrary
integers $N$, and the difference between the savings.}

\label{fig:Ex4_4}
\end{figure}

\begin{figure}[t!]
\centering \scalebox{0.8}{\includegraphics[scale=0.75]{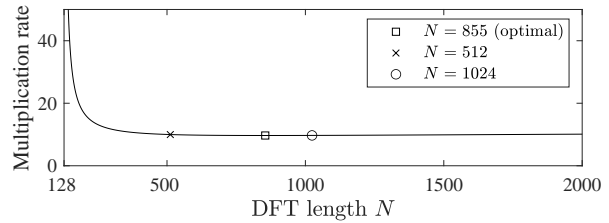}}
\caption{Example 4: Computational complexity versus DFT length $N$ with filter
length $L=128$.}

\label{fig:Ex4_5}
\end{figure}

Further, Fig. \ref{fig:Ex4_3} plots the DFT length $N$ versus the
filter length $L$, both for the case studied above with $N=2^{P}$
and when $N$ can take on all integers. Although the expression used
for the multiplication rates, given by \eqref{eq:M_FD}, holds only
for $N=2^{P}$, the arbitrary-integer-$N$ case is also considered
here for a comparison. As illustrated in Fig. \ref{fig:Ex4_4}, there
is practically no difference between the two cases. In other words,
the use of an arbitrary-integer-$N$ FFT algorithm, with a computational
complexity as in \eqref{eq:M_FD}\footnote{There exist efficient FFT algorithms for values of $N\neq2^{P}$ that
have complexities similar to \eqref{eq:M_FD} \cite{Burrus_2018}.}, will not offer any further complexity reduction as the selection
of the nearest $N$ satisfying $N=2^{P}$ results in practically the
same computational complexity. The reason is that, for a given $L$,
the function $R_{\textnormal{FD}}$ in \eqref{eq:M_FD} is flat over
a large region around the optimal arbitrary-integer-$N$ case. This
is exemplified in Fig. \ref{fig:Ex4_5} for $L=128$. 

\begin{figure}[t!]
\centering \scalebox{0.8}{\includegraphics[scale=0.75]{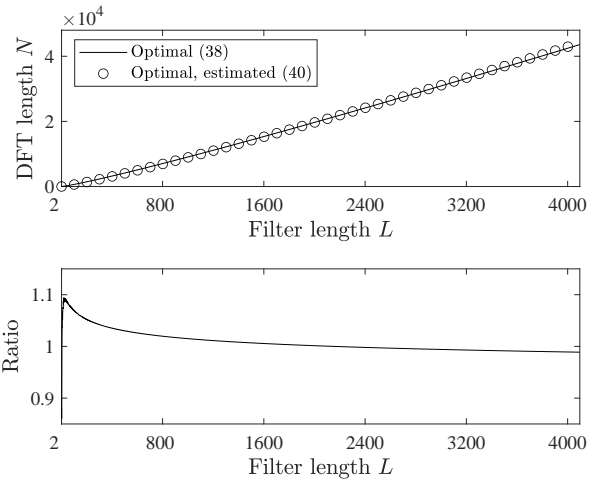}}
\caption{Example 4: Optimal DFT length $N$ and its estimate versus filter
length $L$, and their ratio.}

\label{fig:Ex4_6}
\end{figure}

\subsection{Estimate of the Optimal $N$}

The optimal value of $N$, in the arbitrary-integer-$N$ case, can
be obtained by setting the derivative of $R_{\textnormal{FD}}$ in
\eqref{eq:M_FD} to zero and solve for $N$. This yields
\begin{eqnarray}
N_{\textnormal{opt}} & = & (L-1)\ln(N_{\textnormal{opt}})+C\nonumber \\
 & \approx & (L-1)\ln(N_{\textnormal{opt}}),\quad L>L_{0},\label{eq:Nopt}
\end{eqnarray}
where the constant $C$ is
\begin{eqnarray}
C & = & (1-3\ln(2)/2)(L-1)+4\ln(2)\nonumber \\
 & \approx & -0.03972\times(L-1)+2.773,
\end{eqnarray}
which is much smaller than the term $(L-1)\ln(N_{\textnormal{opt}})$
in \eqref{eq:Nopt} for $L>L_{0}$. For example, with $L_{0}=8$ ($L_{0}=32$),
the ratio between $C$ and $(L-1)\ln(N_{\textnormal{opt}})$ is less
than $10$\% ($1$\%). We have solved equation \eqref{eq:Nopt} numerically
using the Newton-Raphson method with the initial value $N_{\textnormal{opt}}^{(\textnormal{init})}=\hat{N}_{\textnormal{opt}}$,
where the estimated optimal $N$ is

\begin{equation}
\widehat{N}_{\textnormal{opt}}=0.9L\log_{2}(L),
\end{equation}
which is deduced from \eqref{eq:Nopt} and rather close to the optimum
for practical values of $L$. This is illustrated in Fig. \ref{fig:Ex4_6},
where the optimal and estimated optimal values have been rounded to
the nearest integers.

\section{Conclusion\label{sec:Conclusion}}

This paper provided systematic derivations and analyses of MFB and
PTVIR representations of frequency-domain implementations of FIR filters
using the overlap-add and overlap-save techniques. As illustrated
through design examples, including an interpolation example, these
representations are useful when analyzing the effect of coefficient
quantizations as well as the use of shorter DFT lengths than theoretically
required. The examples also illustrated that the PTVIR representation
is preferred when the worst-case time-domain error is more important
than the average error which is captured by the MFB representation.
The paper also provided detailed analysis of the lengths and and relations
between the impulse responses in the PTVIR representation. It was
shown that the overlap-add and overlap-save techniques have different
properties when using quantized coefficients and shorter DFT lengths. 

Finally, a computational-complexity analysis was provided, which showed
that the frequency-domain implementations have lower computational
complexities (multiplication rates) than the corresponding time-domain
implementations for filter lengths that are shorter than reported
earlier in the literature. In particular, for general (unsymmetric)
filters, the frequency-domain implementations turn out to be more
efficient for all filter lengths. For symmetric filters, the frequency-domain
implementations are more efficient for filter lengths of $11$ and
above in the real-signal-and-filter case, and more efficient for odd
(even) filter lengths of $3$ ($6$) and above in the complex-signal-and-filter
case. These results open up for new considerations when comparing
complexities of different filter implementation alternatives.

\bibliographystyle{IEEEtran}
\bibliography{bibliography}

\end{document}